\documentclass[pre,aps,twocolumn,superscriptaddress]{revtex4-2}

\usepackage{graphicx}
\usepackage{booktabs}
\usepackage{amsmath}
\usepackage{amssymb}
\usepackage[utf8]{inputenc}
\usepackage{newunicodechar}
\newunicodechar{‒}{\textendash} 
\usepackage{algorithm}
\floatname{algorithm}{Algorithm}
\usepackage{amsmath,amssymb,amsfonts}
\usepackage{algpseudocode} 
\usepackage{float}           
\usepackage{placeins}        
\usepackage[x11names]{xcolor}    

\usepackage{soul}
\definecolor{lightblue}{rgb}{0.4, 0.7, 0.9}  
\sethlcolor{lightblue}  
\begin{document}

	\title{Scaling law of individual urban tour behavior}
	
	
	\author{Xu-Jie Lin}
	\affiliation{School of Systems Science, Beijing Jiaotong University}
	
	\author{Yitao Yang}
	\affiliation{School of Geography, University of Leeds}
	
	\author{Wei-Peng Nie}
	\email{wpnie@bjtu.edu.cn}
	\affiliation{School of Systems Science, Beijing Jiaotong University}
	
	\author{Xiao-Yong Yan}
	\email{yanxy@bjtu.edu.cn}
	\affiliation{School of Systems Science, Beijing Jiaotong University}
	

	\begin{abstract}
	Analyzing and modeling the mobility process with tour behavior is fundamental to understanding a wide range of complex systems, including animal foraging, human mobility and freight transportation. However, despite their importance, the distribution of tour length has long been neglected in individual human mobility models. To fill this gap, we analyze Foursquare users' check-in data and find that the distribution of urban tour length follows a truncated power-law distribution. To reproduce the universal scaling law for human mobility in urban areas, we introduce a tour terminate–continue model. Our model reproduces not only the urban tour length distribution but also Heaps' law, Zipf's lawand the distribution of the radius of gyration, providing a new perspective for characterizing individual human mobility.
		
	\end{abstract}
	\maketitle
	

	\section{Introduction}
	The mobility process with tour behavior refers to a class of random walks in which an agent explores destinations while intermittently returning to a fixed base location. Such mobility processes appear in diverse domains. For example, animals forage within a limited range and regularly return to their home  \cite{Borger2008}. Human mobility likewise exhibits strong home-return regularities, as individuals frequently return to their residence despite complex daily movements \cite{Belik2011}. Freight vehicles also repeatedly return to a fixed depot during the freight transportation process \cite{14}. These examples show that mobility processes with tour behavior emerge broadly in natural and social systems.

	Among these domains, analyzing and modeling human mobility patterns has become increasingly important in a wide range of applications, such as urban planning, transportation engineering, and public health \cite{3}. Empirical studies have revealed many fundamental statistical characteristics and universal scaling laws of individual human mobility \cite{23}. For example, the trip distance distribution and the distribution of the radius of gyration follow truncated power-law \cite{7}, location visit frequencies obey Zipf’s law, and the distinct number of locations visited over time follows Heaps' law \cite{8}. However, these studies overlook the scaling law of individual tour behavior, which is critical for capturing the complex pattern of human mobility. A tour is a chain of trips that starts and ends at a base location \cite{10,11,9}. For example, human tours tend to include one or two intermediate stops around work or school for commuting or shopping \cite{12}. Analyzing and modeling the statistical properties of individual tour behavior can provide valuable insights into the mechanisms of human mobility and inform more effective transportation planning and management strategies.
	
	Most recent studies that analyze tour behavior have focused on quantifying tour characteristics \cite{11,14,15}, classifying tour behavior \cite{11,15,16}, analyzing daily tour patterns \cite{17,9} and investigating the relationship between tour behavior and mode choice \cite{18}. However, they do not  consider the distribution of tour length (the number of intermediate stops in a tour), which reflects the overall tour choice behavior. Some studies have conducted preliminary statistical analyses of tour length of human mobility \cite{19}. Schneider et al. reported that, in both Paris and Chicago, the probability of an individual mobility motif decreases with its stop number \cite{19}. Although the motif distribution can partially reflect the fact that the probability of a tour terminating decreases as the tour length increases, different kinds of motifs may have the same tour length. Consequently, there is still a lack of research that examines the distribution of tour length and why it follows this distribution.

	Although there is currently no model that reproduces tour length distributions, a variety of models have been developed to describe individual human mobility behavior, e.g., \cite{7, 8, 23, 24, 25, 26, 27, 28, 29, 86,87,88,89,90}. These models provide us with valuable insights. Early models use Levy random walks to model human mobility and accurately replicate the observed statistical characteristics \cite{24,25}, but they neglect some characteristics, such as individuals’ preferences for returning to previously visited locations \cite{7,8}. In response to this problem, Song et al. developed an exploration and preferential return (EPR) model by considering individuals’ tendency to revisit locations and explore new locations \cite{7}. The EPR model can successfully reproduce statistical characteristics such as Zipf's law \cite{26} and Heaps' law \cite{27} of human mobility. Yan et al. proposed a simplified human mobility universal (SHMU) model that considers individuals’ memory effects when choosing the next location \cite{28}. This model uses only one parameter to characterize individuals’ memory of locations and reproduce the statistical characteristics of human mobility. While these models can reproduce the statistical characteristics of human mobility using simple rules, they do not consider the tour generation mechanism. Lenormand et al. proposed a new model that considered individuals’ choice of terminating the search process of shopping and successfully reproduced the distribution of shopping trip distance \cite{29}. Although it was not originally designed to model individual tour behavior, its framework highlights the critical role of the tour termination choice process in human mobility.

	In transportation science, several studies that focus on modeling the tour termination choice process provide valuable insights. Hunt and Stefan proposed a tour generation framework using a discrete choice model to predict the purpose of the next stop of commercial vehicles \cite{13}. Furthermore, Wang and Holgu\'{i}n-Veras proposed a simplified tour termination model using a binary logit model to decide whether an individual will return to the base location or continue the current tour \cite{30}. However, these models require many parameters, which not only increases model complexity but also causes the model to fail to capture the core factors that influence tour choice behavior. Although individual tour behavior is the result of joint interactions across multiple decision-makers and environments \cite{18,31,9}, the core factors that influence individual tour behavior must be captured. In contrast, studies that model human mobility using statistical physics approaches tend to use simple rules and few parameters to reproduce scaling laws. Therefore, integrating the tour generation framework from transportation science with individual random walk models from statistical physics is a promising research direction for uncovering the underlying mechanisms behind the tour length distribution.
	
	In this study, we first uncover the urban tour length distribution and then develop a model to reproduce the statistical characteristics of individual human mobility, including the tour length distribution, Heaps’ law, Zipf’s law and the distribution of the radius of gyration. We start by presenting the Foursquare users' check-in dataset used for empirical analysis. This dataset contains information on the trajectories of Foursquare users, which represent intracity human mobility. We then introduce a model that considers the process by which an individual decides to terminate a tour and individuals’ tendency to revisit locations. Additionally, we extend the model by a gravity-based mechanism that governs the exploration of new locations within the urban area. The model generates the tour length distribution, Heaps’ law, Zipf’s law and the distribution of the radius of gyration, which mimic the empirical results and can explain the underlying mechanism of the tour length distribution for human.

	\section{Data}
	We use Foursquare users' check-in data in our work, which consists of Foursquare users' check-in records in New York and Los Angeles, including 255,204 trips generated by 15,499 users \cite{32}. Foursquare users' check-in data directly reflect voluntary visits to locations, making these data particularly valuable for analyzing individual human mobility. Each record in these two cities contains an individual ID, timestamp, and geographic coordinates. We identify the most frequently visited location for each individual as the base location and extract tours from the trajectory data. We filter out individuals with fewer than two trips. Table \ref{tab. 1} presents the final number of individuals, number of trips, number of tours, mean tour length, variance of tour length in the two cities.

	\begin{table}[h!]
		\centering
		\caption{Number of individuals, number of trips, number of tours, mean tour length, and variance of tour length in New York and Los Angeles}
		\label{tab. 1}
		\begin{tabular}{lrr}
			\toprule
			\textbf{Metric} & \textbf{New York} & \textbf{\quad Los Angeles} \\
			\midrule
			\textbf{Number of individuals} & 9{,}948 & 5{,}551 \\
			\textbf{Number of trips}       & 155{,}314 & 99{,}890  \\
			\textbf{Number of tours}       & 20{,}306 & 9{,}743  \\
			\textbf{Mean tour length}       & 7.42 & 6.02 \\
			\textbf{Variance of tour length}       & 158.40 & 92.40 \\
			\bottomrule
		\end{tabular}
	\end{table}
	
	\section{Empirical analysis}
	\begin{figure}[h]
		\centering
		\includegraphics[width=0.5\textwidth]{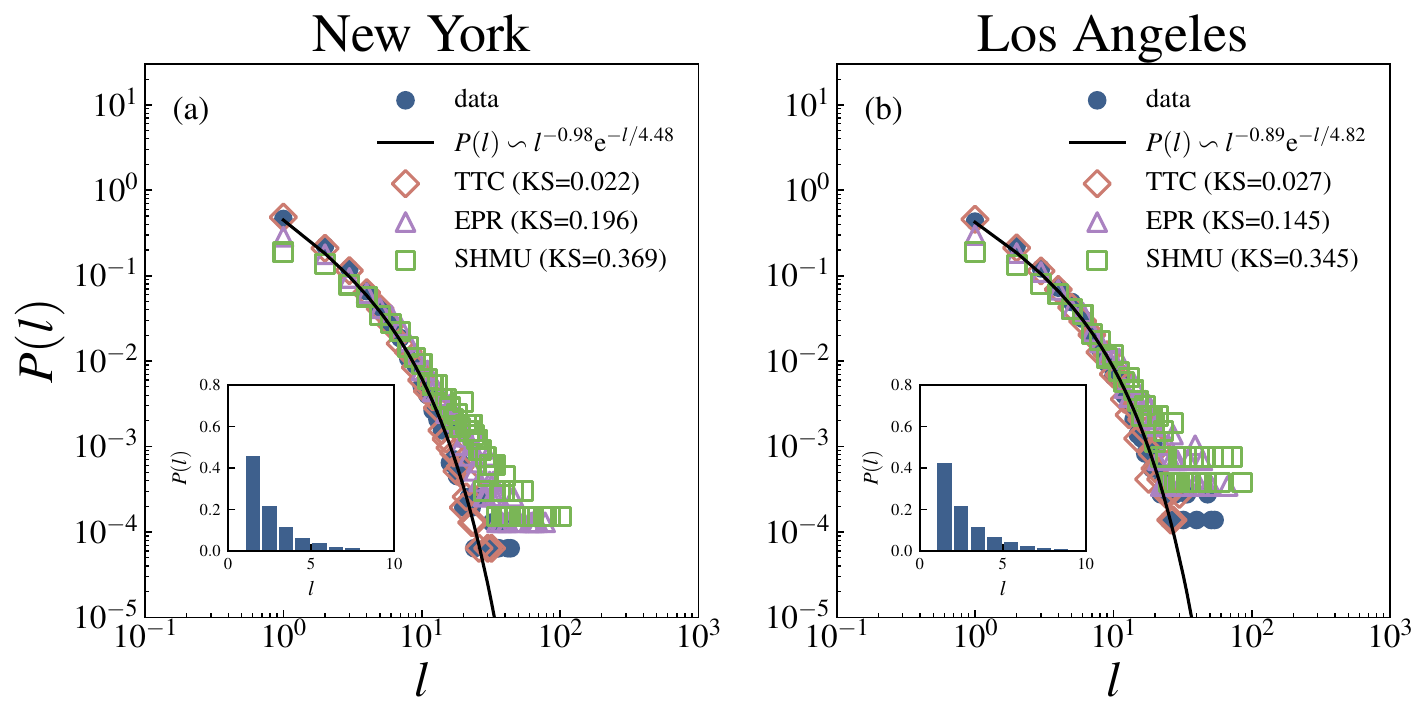} 
		\caption{Tour length distributions $P(l)$ of New York (a) and Los Angeles (b), where $P(l)$ denotes the proportion of tours with length $l$ relative to the total number of tours. The main figure shows the distribution of tour lengths in log-log coordinates, where the blue dots represent the empirical data and the black curve represents the fitted values. The diamonds, triangles and squares represent the outputs of the TTC, EPR and SHMU models, respectively. The inset bar chart displays the tour length distribution using linear coordinates. Statistics. Goodness of fit is assessed by the Kolmogorov–Smirnov (KS) distance between the empirical and model distributions; values are given in the legend.}
		\label{fig. 1}
	\end{figure}
	We analyze the tour length distributions in two cities, as shown in Fig. \ref{fig. 1}. The tour length distributions of human mobility are well approximated by a truncated power-law
	\begin{equation}
		P(l)\sim {{l}^{-\alpha }}{{\mathrm{e}}^{-l/\beta }},
		\label{eq. 1}
	\end{equation}
	where $l$ is the tour length, $\alpha$ is the scaling exponent and $\beta$ is the cutoff value(Note that $l$ is in distinction to $n$, which represents the number of trips of an individual). The observed truncated power-law distribution captures the heterogeneity in the individuals' choice of tour length, i.e., most individuals return to the base location after visiting only one or two intermediate stops, whereas a minority of individuals visit many intermediate stops before returning. The exponential tail effect of the tour length distribution indicates that while short- and medium-length tours are frequent, the probability of observing very long tours diminishes rapidly because of city size limitations or daily scheduling constraints \cite{7}.

	\section{Model}
	The tour length distributions of human mobility within U.S. cities universally follow a truncated power-law. However, why do individuals in different environments exhibit similar patterns? There may be a common mechanism that governs the generation of tours for human mobility in different cities. To our knowledge, the distribution of tour length has received limited attention in the study of individual human mobility. We adopt the EPR and SHMU models to investigate the ability of traditional human mobility models to reproduce tour length distributions. Figure \ref{fig. 1}  shows that the EPR model and SHMU model cannot reproduce the empirically observed distribution of tour lengths. This limitation arises from the fact that the EPR model primarily captures individuals’ exploration of new locations and preferential return to previously visited locations, whereas the SHMU model focuses solely on the memory effect associated with location visit frequency. Crucially, both models neglect the tour generation processes. Although tours can be extracted from the simulated trajectories, the tour length distributions generated by both models exhibit substantial discrepancies compared with empirical observations. To fill this gap, the key factors that influence individuals’ decisions to terminate a tour must be captured.

	\begin{figure*}[t]
		\centering
		\includegraphics[width=0.9\textwidth]{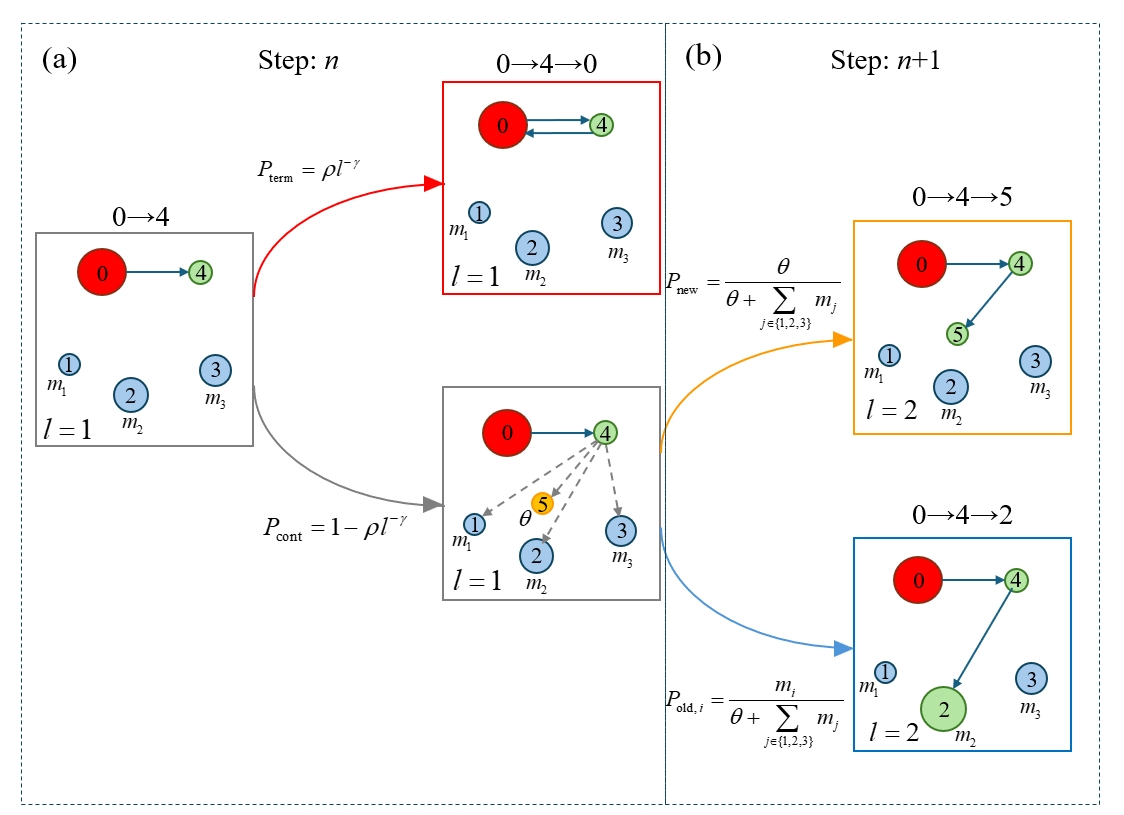} 
		\caption{A schematic diagram of the TTC model. At each step, the individual first departs from the current location and decides whether to terminate the tour and return to the base location (red circle) with a probability $P_{\text{term}}$ related to the current tour length $l$. If the individual chooses to continue the tour, they then decide whether to visit a new location (yellow circle) or a previously visited location (blue circle), according to the competitive mechanism involving parameter $\theta$ and the location visit frequency $m_{k}$.}
		\label{fig. 2}
	\end{figure*}
	\begin{figure}[h]
		\centering
		\includegraphics[width=0.5\textwidth]{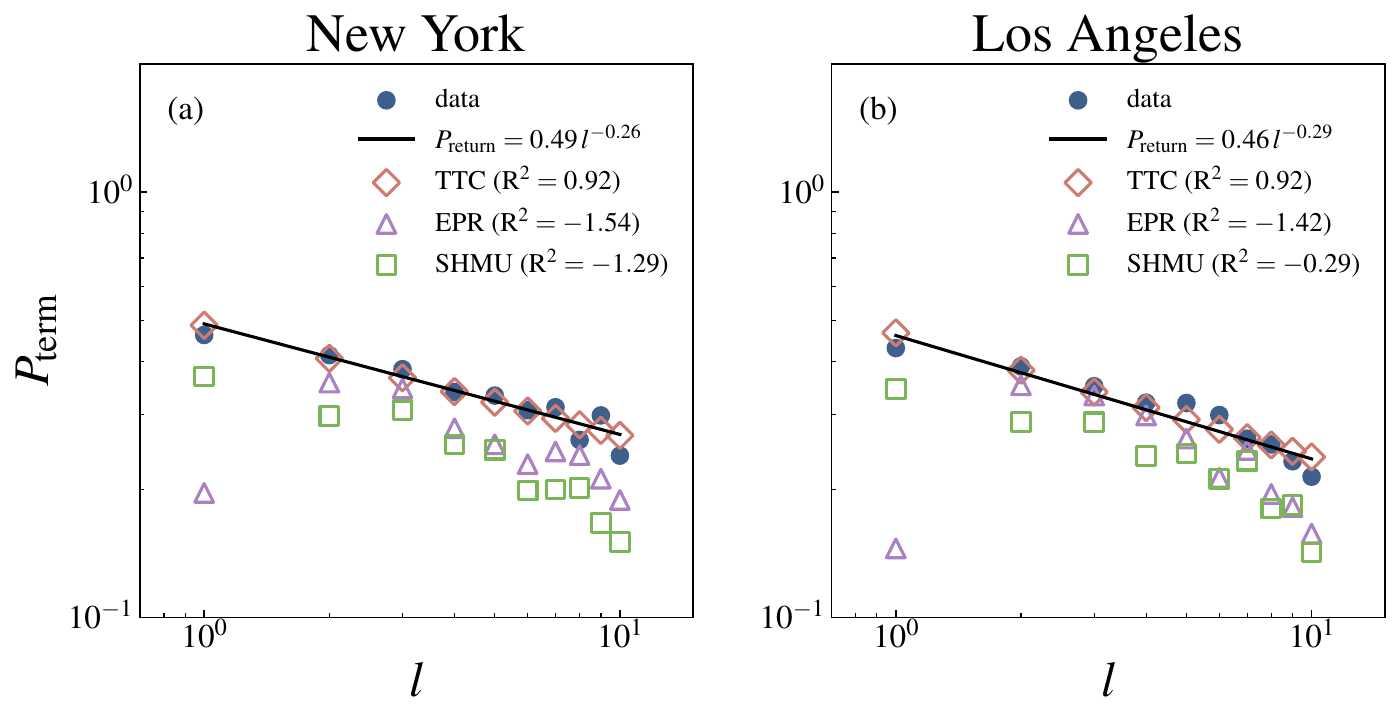} 
		\caption{Scaling relationships between the probability $P_{\text{term}}$ of terminating the tour and the length $l$ of the current tour in New York (a) and Los Angeles (b), where $P_{\text{term}}(l)$ indicates the probability of terminating a tour with length $l$. Note that the denominator of this probability is the number of tours with length $\geqslant l$. The blue dots represent the empirical data, whereas the black line represents the linear regression fit in the log-log coordinates. The diamonds, triangles and squares represent the outputs of the TTC, EPR and SHMU models, respectively. Statistics. Goodness of fit is quantified by $\mathrm{R^{2}}$ computed in linear space; values are given in the legend.}
		\label{fig. 3}
	\end{figure}
	\begin{figure}[h]
		\centering
		\includegraphics[width=0.5\textwidth]{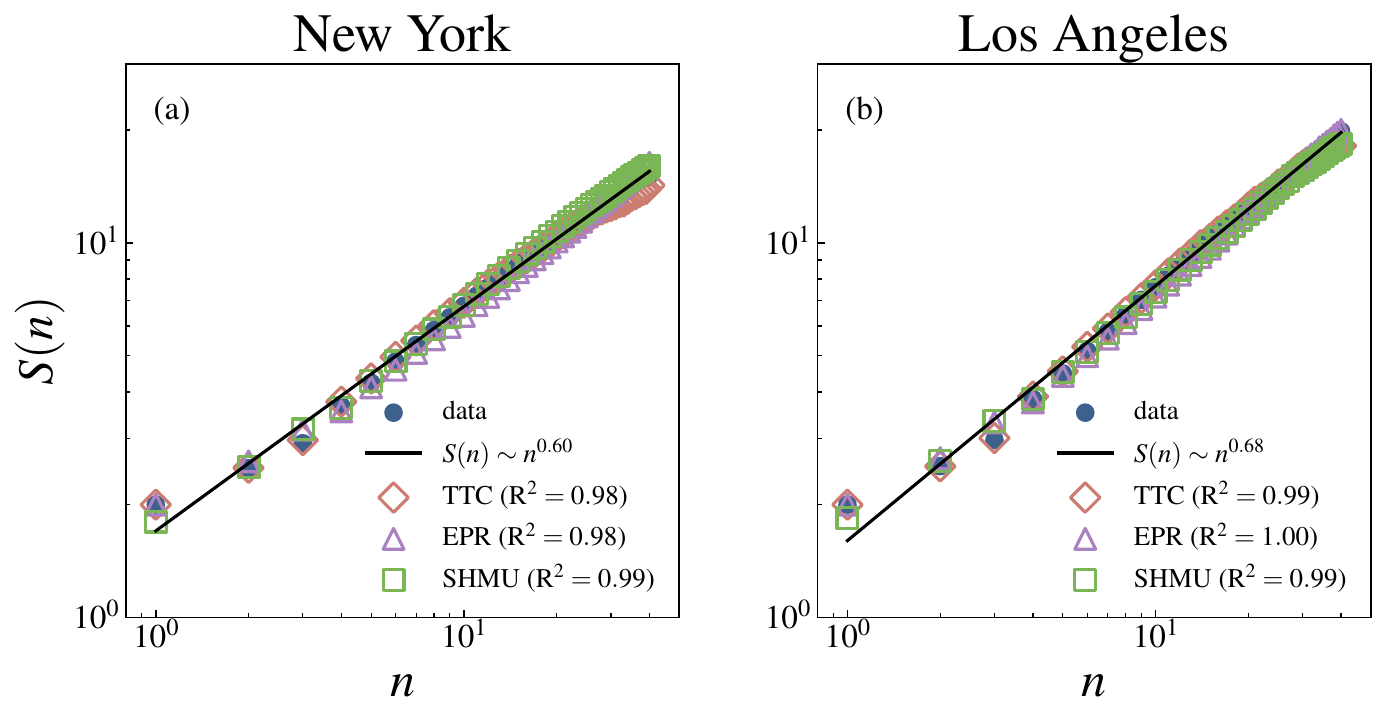} 
		\caption{Relationship between the number $S$ of locations visited and the number $n$ of trips in New York (a) and Los Angeles (b), where $S(n)$ denotes the total number of distinct locations visited after $n$ trips. Statistics. Statistical measures and notations follow those of Fig. \ref{fig. 3}. }
		\label{fig. 4}
	\end{figure}
	
	We propose a tour terminate-continue (TTC) model, as shown in Fig. \ref{fig. 2}. In the TTC model, we assume that an individual begins his or her tour at a base location. At each step, the individual must decide whether to terminate the tour and return to the base location (see Fig. \ref{fig. 2}(a)). This tour generation process is modeled through the relationship between the probability of terminating the tour and the length of the current tour. Figure \ref{fig. 3} shows the empirical relationship between the probability of terminating the tour and the length of the current tour in the two cities. We find that the probability of terminating the tour decreases with increasing current tour length, which is consistent with the findings of Hunt and Stefan \cite{13}. Moreover, we observe a negatively correlated scaling relationship between the probability ${P}_{\text{term}}$ of terminating a tour and the current tour length $l$, i.e.,
	\begin{equation}
		{{P}_{\text{term}}}=\rho {{l}^{-\gamma },}
		\label{eq. 2}
	\end{equation}
	where $\rho$ denotes the initial probability of termination, and $\gamma$ characterizes how the probability of termination decreases as the tour increases in length. Accordingly, the probability of continuing the tour can be expressed as
	\begin{equation}
		{{P}_{\text{cont}}}=1-\rho {{l}^{-\gamma }}.
		\label{eq. 3}
	\end{equation}
	The parameter $\rho$ represents the initial probability of an individual choosing to terminate the tour. The greater the value of $\rho$ is, the higher the probability $P_{\text{term}}$ of  termination. The parameter $\gamma$ determines how strongly the current tour length $l$ influences the probability of termination. When $\gamma=0$, the probability of termination is independent of the current tour length $l$. When $\gamma>0$, the probability of termination decreases with the increasing current tour length $l$, following a power-law form. 
	
	When an individual chooses to continue the current tour, it is necessary to determine whether the next step involves visiting a new location or a previously visited location (see Fig. \ref{fig. 2}(b)). To capture the individual's preference for returning to previously visited locations, we introduce a parameter $\theta$ that characterizes the competition between visiting new and old locations. We assume  that the probability of an individual exploring a new location is	
	\begin{equation}
		{{P}_{\text{new}}}=\frac{\theta }{\theta +\sum\limits_{\substack{k>0 \\ k \neq k'}}{{{m}_{k}}}},
		\label{eq. 4}
	\end{equation}
	where $\theta$ is the exploration parameter that characterizes an individual's tendency to explore a new location, ${m}_{k}$ is the number of visits to location $k$, $k = 0$ represents the index of the base location, and $k'$ represents the index of the last visited location. Note that the new location must differ from both the base location and the last visited location. This implies that both $k = 0$ and $k = k'$ are excluded from the summation in Eq. (\ref{eq. 4}). Accordingly, the probability of an individual revisiting an old location $k$ (not the base location) is
	\begin{equation}
		{{P}_{\text{old, }k}}=\frac{{{m}_{k}}}{\theta +\sum\limits_{\substack{k>0 \\ k \neq k'}}{{{m}_{k}}}}.
		\label{eq. 5}
	\end{equation}
	After the next location to visit is selected, the process returns to Fig.  \ref{fig. 2}(a). Eq. (\ref{eq. 4}) and Eq. (\ref{eq. 5}) indicate that a larger value of $\theta$ indicates a greater tendency for individuals to explore new locations. When the exploration parameter $\theta=0$, the individual will not explore any new locations but will select the next destination among previously visited locations; the probability is proportional to the individual’s visit frequency. When the exploration parameter $\theta$ approaches infinity, the individual will explore only new locations rather than returning to previously visited nonbase locations.

	\begin{figure}[h]
		\centering
		\includegraphics[width=0.5\textwidth]{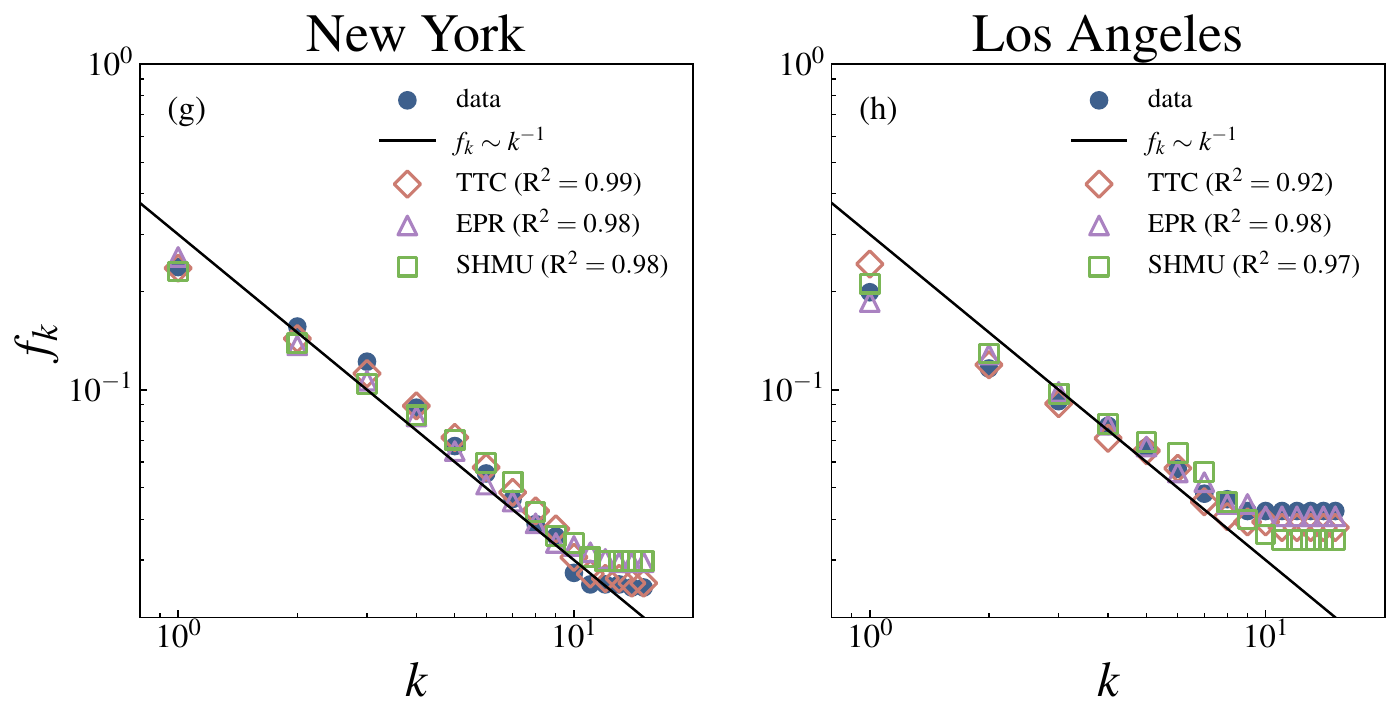} 
		\caption{Distributions of the frequency of visits $f_{k}$ to locations in New York (a) and Los Angeles (b),  where $k$ is the rank of locations and $f_{k}$ represents the frequency of visiting location $k$. The blue dots represent the empirical data, whereas the black line represents the standard Zipf's law $f_{k}\sim k^{-1}$. The diamonds, triangles, and squares represent the outputs of the TTC, EPR and SHMU models, respectively. Statistics. Statistical measures follow Fig. \ref{fig. 3}.
		}
		\label{fig. 5}
	\end{figure}
	
	\section{Results}
	\subsection{Model calibration}
	The TTC model has three parameters $\rho$, $\gamma$ and $\theta$. The parameters $\rho$ and $\gamma$, which are estimated from empirical data (see Fig. \ref{fig. 3}), control the probability of terminating the current tour. Parameter $\theta$ influences individuals' tendency to explore new locations and is the only parameter that requires calibration. The model and data are compared according to the relationship between the number $S$ of locations visited and the number $n$ of trips. We adjusted the value of parameter $\theta$ by minimizing the relative error between the observed ${S}_{\text{data}}$ and the simulated result ${S}_{\text{model}}$
	\begin{equation}
		E=\sum\limits_{n=1}^{{{n}_{\max }}}{\frac{\left| {{S}_{\text{data}}}(n)-{{S}_{\text{model}}}(n) \right|}{{{S}_{\text{data}}}(n)}}.
		\label{eq. 6}
	\end{equation}
	Using Eq. (\ref{eq. 6}), we can obtain the optimal values of parameter $\theta$ in the two cities.
	
	\subsection{Model prediction}
	
	We focus on the following statistical properties of urban tour-based human mobility: (1) the distribution of tour length, (2) the relationship between the probability of termination and the current tour length, (3) the relationship between the number of locations visited and the number of trips, and (4) the frequency of location visits. To validate the effectiveness of the proposed model, we select two classical individual human mobility models, namely, the EPR and SHMU models, as benchmarks for comparison. The parameters of the benchmark models are calibrated using the method described in Eq. (\ref{eq. 6}).
	
	Figure \ref{fig. 1} presents the simulation results of the tour length distribution for the TTC, EPR, and SHMU models across different cities.  We compared models by their Kolmogorov–Smirnov statistic KS computed between empirical data and simulation results (lower is better). There is excellent agreement between the results of the TTC model and the empirical results. In the TTC model, the decrease in tour frequency with increasing tour length is attributed to the regularity in individuals’ decisions to terminate their tours and return to the base location (see Eq. (\ref{eq. 2})). Nonetheless, the EPR model and SHMU model yield markedly larger KS distances than TTC model, as they underestimate the proportion of short tours and overestimate the proportion of long tours in comparison with the empirical data. This suggests that while tours can be extracted from the output trajectory of both models, both EPR and SHMU model fail to reproduce the tour length distributions because of their inability to capture the actual tour generation process.
	
	Figure \ref{fig. 3} presents the simulation results of the relationship between the probability $P_{\text{term}}$ of termination and the current tour length $l$ for different models in different cities, as well as with the empirical results for comparison. We quantify model fit using the goodness of fit ($\mathrm{R^{2}}$). The results demonstrate that the outputs of the TTC model are the closest to the empirical data because the tour generation mechanism of the TTC model is in accordance with the empirical scaling relationship in Eq. (\ref{eq. 2}). In contrast, the EPR model and SHMU model underestimate the probability of terminating the tour. In these two models, individuals do not choose to return to a specific base location but tend to visit locations with higher visit frequencies. This discrepancy arises because both models consider only the individual's preference for revisiting previously visited locations, without incorporating the actual tour generation mechanism.
	
	Figure \ref{fig. 4} shows the simulation results of the relationship between the number of locations visited and the number of trips for different models in the two cities. We quantify model fit using the goodness of fit ($R^{2}$). The simulated results of all three models agree with the empirical data and approximately follow Heaps’ law. This suggests that all the models can capture individuals’ preferences for previously visited locations.
	
	Figure \ref{fig. 5} shows the simulation results of location visit frequencies for different models in the two cities. We quantify model fit using the goodness of fit ($\mathrm{R^{2}}$). The simulated results of all three models agree with the empirical data and approximately follow Zipf’s law. This finding suggests that all the models successfully capture the heterogeneity in the attractiveness of locations.
	
	\subsection{Spatial dimension analysis}
	Although the TTC model can successfully reproduce tour length distribution, Zipf's law, and Heaps' law, a key limitation is the lack of spatial dimension: all locations are treated as equally accessible regardless of distance. In practice, tour generation is influenced by spatial factors, such as distance, spatial constraints and location attractiveness \cite{Kapitza2025, PangZhang2019, ChowdhuryScott2020}. For example, individuals consider travel distance in tour generation to enhance commuting efficiency and tend to prefer locations with more population to seek opportunities. Omitting this information can lead to the TTC model's failure to reproduce the spatial characteristics of tour-based human mobility, such as the distribution of the radius of gyration \cite{dEPR2016}. Inspired by the d-EPR model \cite{dEPR2016, 37}, we augment the exploration step of the TTC model by incorporating destination attractiveness and trip distance into the probability $P_{new}^{(j)}$ of choosing the new location $j$, which is
	\begin{equation}
		P_{new}^{(j)} \sim \frac{O_{i}O_{j}}{d_{i j}^{\sigma}},
		\label{eq. 12}
	\end{equation}
	where $O_{i}$ is the population of the current location $i$, $O_{j}$ is the population of the next location $j$, $d_{i j}$ is the distance between the current location $i$ and the next location $j$ and the parameter $\sigma$ captures how strongly distance affects the probability $P_{new}^{(j)}$. We calibrated the parameter by minimizing the Kolmogorov–Smirnov distance between the empirical distribution and the simulated one. We refer to the modified model as the d-TTC model (see Appendix A, Algorithm 4).
	\begin{figure}[h]
	\centering
	\includegraphics[width=0.5\textwidth]{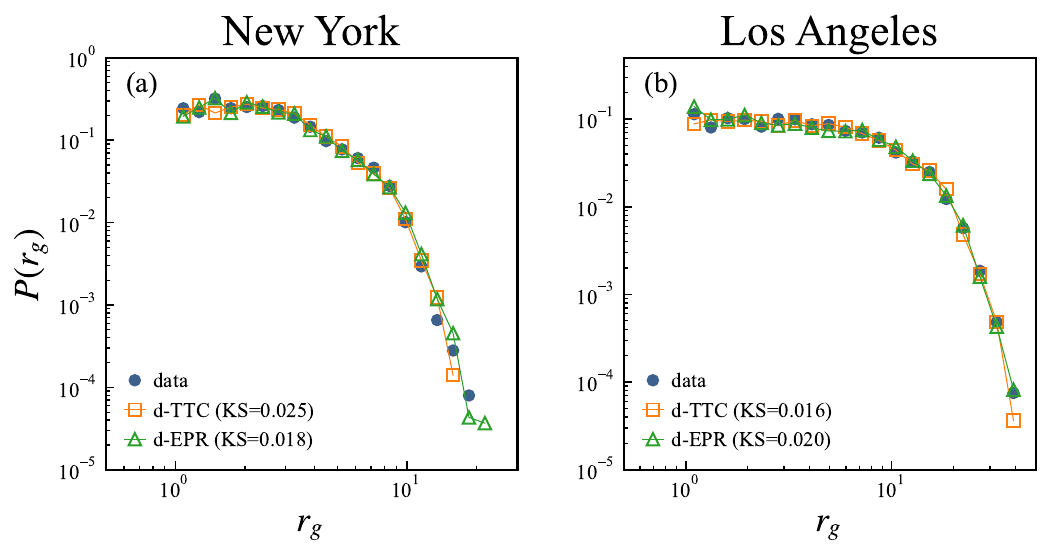} 
	\caption{Distributions $P(r_{g})$ of the radius of gyration in New York (a) and Los Angeles (b),  where $r_{g}$ is the radius of gyration of an individual. The blue dots represent the empirical data. The squares and triangles represent the outputs of the d-TTC and d-EPR models, respectively. Statistics. Statistical measures follow Fig.  \ref{fig. 1}. }
	\label{fig. 6}
	\end{figure}
	
	Figure \ref{fig. 6} presents the simulation result of the distribution of the radius of gyration $r_{g}$ of individual trajectory for the d-TTC model and the d-EPR model. Both models show excellent agreement between simulation results and the empirical results. These results show that the d-TTC model is able to reproduce the empirical radius of gyration distribution. Note that this comparison is not a performance benchmark; rather, it demonstrates the TTC model’s extensibility in the spatial dimension.
	\subsection{Analytical results}
	According to the TTC model, the probability that a tour has a length of $l$ can be formulated as
	\begin{equation}
		P(l)={{P}_{\text{term}}}(l)\prod\limits_{i=1}^{l-1}{{{P}_{\text{cont}}}(i)}.
		\label{eq. 7}
	\end{equation}
	By substituting Eq. (\ref{eq. 2}) into (\ref{eq. 7}), we can obtain the theoretical distribution of tour length
	\begin{equation}
		P(l)=\rho {{l}^{-\gamma }}\prod\limits_{i=1}^{l-1}{(1-\rho {{i}^{-\gamma }})}.
		\label{eq. 8}
	\end{equation}
	Furthermore, on the basis of Eq. (\ref{eq. 2}) and Eq. (\ref{eq. 4}), the average changes in the number $S$ of locations at step $n$ are determined by the location attributes at step $n-1$ and the choice behavior at step $n$. If the base location is visited at step $n-1$, then a nonbase location must be visited at step $n$; If a nonbase location is visited at step $n-1$, then the location visited at step $n$ can be either the base or a nonbase location. Therefore, the average changes in the number $S$ of locations at step $n$ can be approximated by a differential equation
	
	\begin{equation}
		\frac{\text{d}S}{\text{d}n}={P}^{(n-1)}_{\text{cont}}{P}_{\text{cont}}^{(n)}\frac{\theta }{n+\theta }+{P}^{(n-1)}_{\text{term}}\frac{\theta }{n+\theta },
		\label{eq. 9}
	\end{equation}
	where ${P}_{\text{cont}}^{(n)}$ and ${P}_{\text{term}}^{(n)}$ represents the probability of continuing and terminating the tour at step $n$ respectively.	By approximating ${P}_{\text{term}}^{(n)}(\rho, \gamma)$ with its mean value $\left\langle {{P}_{\text{term}}} \right\rangle (\rho, \gamma)$, we obtain
	\begin{equation}
		\frac{\text{d}S}{\text{d}n}=\left[ {{(1-\left\langle {{P}_{\text{term}}} \right\rangle (\rho, \gamma) )}^{2}}+\left\langle {{P}_{\text{term}}} \right\rangle (\rho, \gamma) \right]\frac{\theta }{n+\theta }.
		\label{eq. 10}
	\end{equation}
	The solution to Eq. (\ref{eq. 10}) is simply
	\begin{equation}
		S=\left[ 1-\left\langle {{P}_{\text{term}}} \right\rangle (\rho, \gamma)+{{\left\langle {{P}_{\text{term}}} \right\rangle (\rho, \gamma)}^{2}} \right]\theta \ln (1+\frac{n}{\theta })+C,
		\label{eq. 11}
	\end{equation}
	where $C$ is a constant.
	
	To validate the analytical results and investigate how different parameters influence the model output, we perform a series of simulations. Taking Foursquare users' check-in data from New York as an example, we first calibrate the optimal parameters $\rho$, $\gamma$, $\theta$ and $\sigma$ for the TTC model, and then analyze the simulation and theoretical results by changing a single parameter while keeping the others unchanged.
	\begin{figure}[h]
		\centering
		\includegraphics[width=0.48\textwidth]{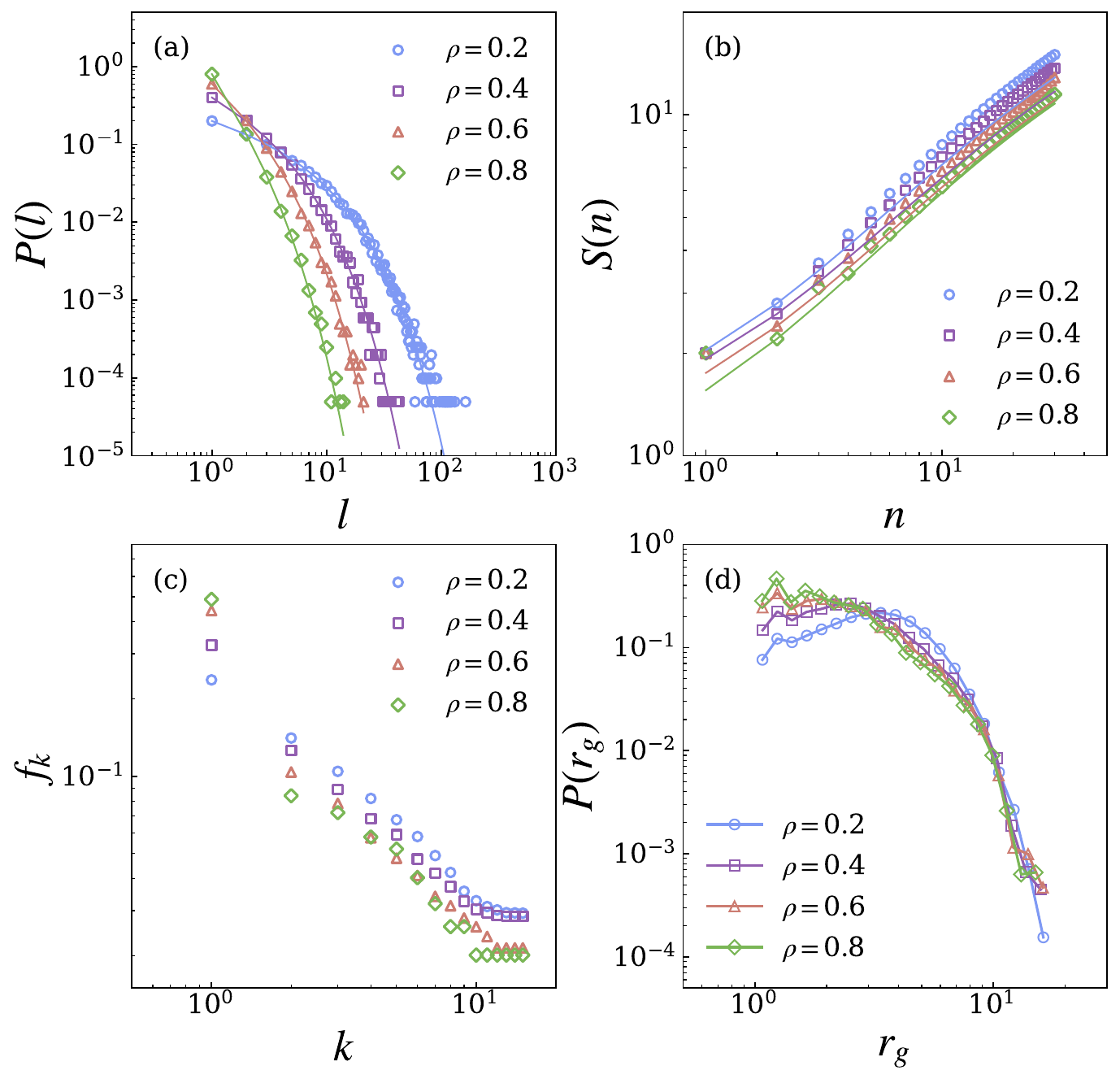}
		\caption{Effects of the parameter $\rho$ on the simulation (scatter points) and analytical (curves) results. The parameter $\rho$ is set to 0.2 (circle), 0.4 (square), 0.6 (triangle), and 0.8 (diamond). (a) Tour length distribution $P(l)$. (b) Relationship between the number $S$ of locations visited and the number $n$ of trips. (c) Frequency $f_{k}$ at which a location $k$ is visited. (d) The distribution $P(r_{g})$ of radius of gyration.}
		\label{fig. 7} 
	\end{figure}
	
	Figure \ref{fig. 7} presents the TTC model’s simulation and analytical results as parameter $\rho$ increases from 0.2 to 0.8. As shown in Fig. \ref{fig. 7}(a), the analytical (see Eq. (\ref{eq. 8})) and simulation results show good agreement in terms of the tour length distribution. Moreover, the probability of longer tours decreases, whereas the probability of shorter tours increases, indicating that individuals often undertake shorter tours within the city. Consequently, as illustrated in Fig. \ref{fig. 7}(b), the growth rate of $S$ decreases as $\rho$ increases since individuals are more likely to return to the base location than to visit new locations, reflecting a reduction in individuals' tendency toward exploration. Moreover, the analytical (see Eq. (\ref{eq. 11})) and simulation results show good agreement in terms of the increase trend in the number $S$ of locations visited. Additionally, Fig. \ref{fig. 7}(c) demonstrates that the frequency at which an individual visits the base location increases, whereas the frequency of visits to nonbase locations decreases with increasing parameter $\rho$, suggesting that greater values of parameter $\rho$ increase the heterogeneity of the location visit frequency distribution. Finally, Fig. \ref{fig. 7}(d) shows that, as the parameter $\rho$ increases, the probability of a low radius of gyration $r_{g}$ increases. This result is obtained because the decrease of tour length leads to the concentration of the trajectory near the base location and thus reducing the overall radius of gyration $r_{g}$.
	\begin{figure}[h]
		\centering
		\includegraphics[width=0.48\textwidth]{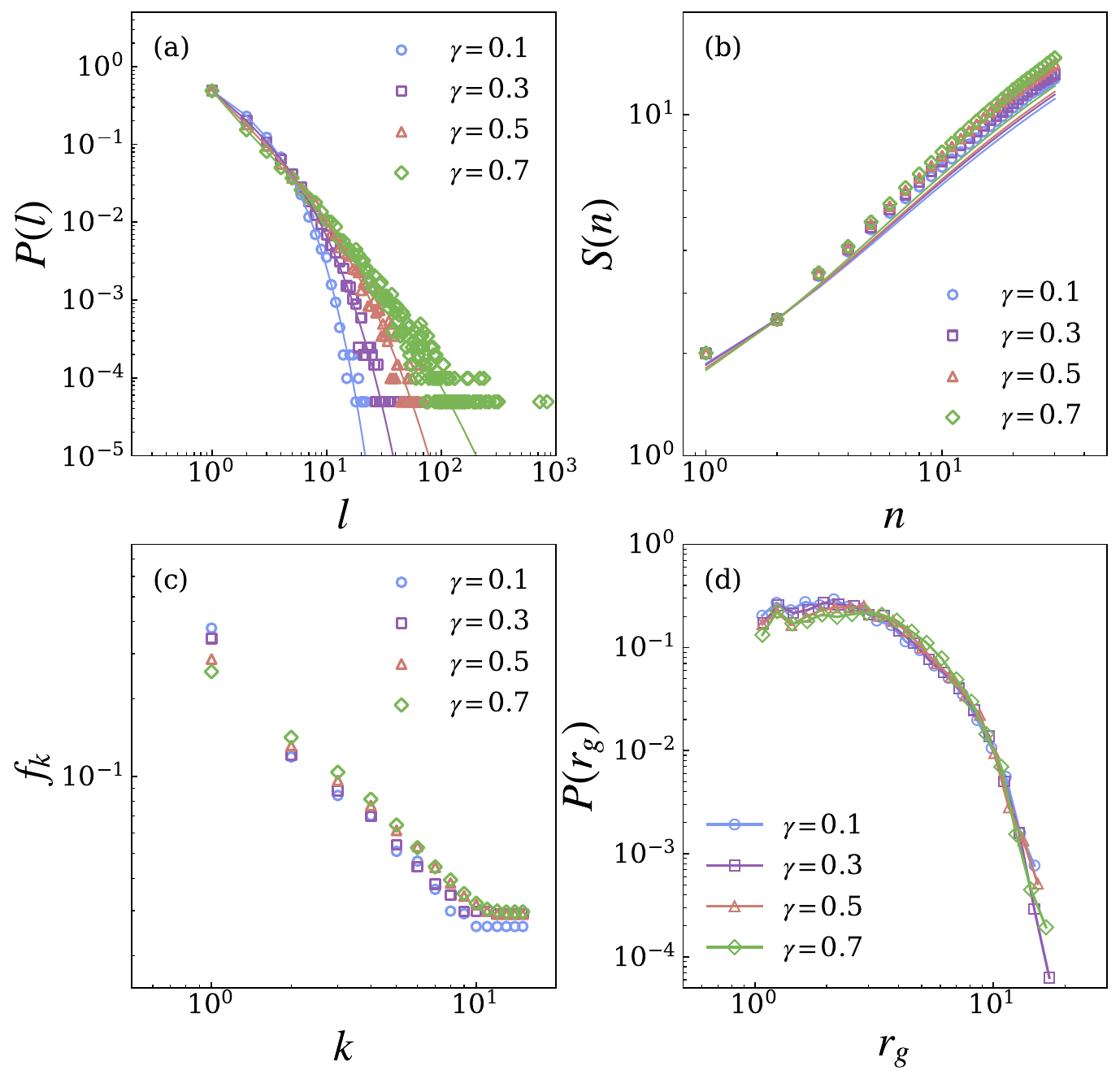}
		\caption{Effects of parameter $\gamma$ on the simulation (scatter points) and analytical (curves) results. The parameter $\gamma$ is set to 0.1 (circle), 0.3 (square), 0.5 (triangle), and 0.8 (diamond). (a)-(d) have the same meaning as do those in Fig. \ref{fig. 7}.}
		\label{fig. 8}
	\end{figure}
	
	Figure \ref{fig. 8} presents the simulation and analytical results as parameter $\gamma$ increases from 0.1 to 0.7. As shown in Fig. \ref{fig. 8}(a),  the analytical and simulation results show good agreement in terms of the tour length distribution. Moreover, the probability of long tours decreases, whereas the probability of short tours slightly increases, indicating that individuals often undertake shorter tours within the city and are more likely to terminate their tours. Consequently, as illustrated in Fig. \ref{fig. 8}(b), the growth rate of $S$ slightly increases as $\gamma$ increases, since individuals are less likely to return to the base location and more likely to visit new locations, reflecting an increase in individuals' tendency toward exploration. Moreover, the analytical and simulation results show good agreement in terms of the increase trend in the number $S$ of locations visited. Additionally, Fig. \ref{fig. 8}(c) shows that the frequency at which individuals visit base locations decreases, whereas the frequency of visits to nonbase locations increases, suggesting that greater values of parameter $\gamma$ reduce the heterogeneity of the location visit frequency distribution. Finally, Fig. \ref{fig. 8}(d) shows that, as the parameter $\gamma$ increases, the distribution $P_{r_{g}}$ of the radius of gyration remains nearly unchanged, indicating that the parameter $\gamma$ has a negligible effect on $r_{g}$.
	\begin{figure}[h]
		\centering
		\includegraphics[width=0.48\textwidth]{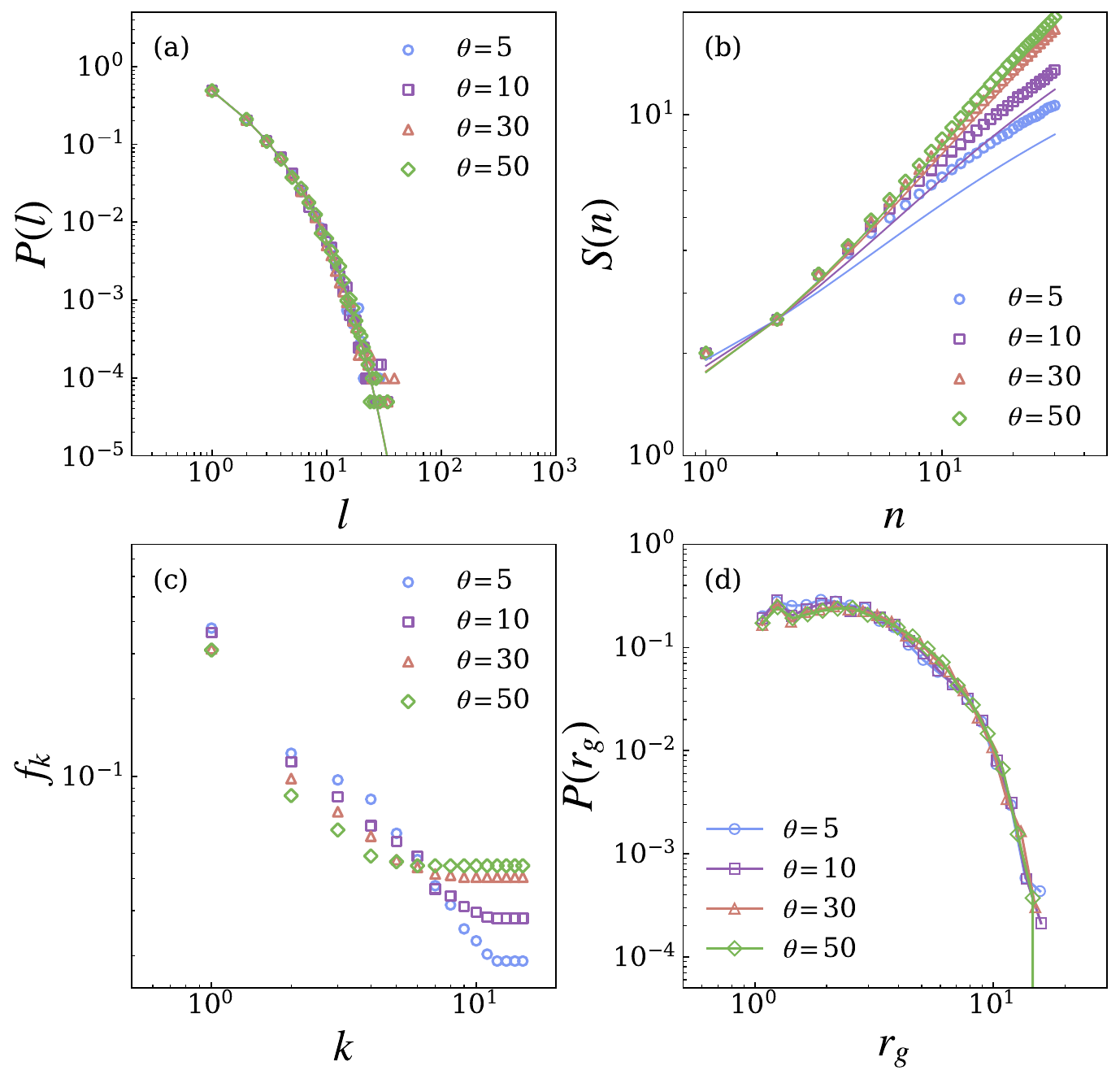}
		\caption{Effect of parameter $\theta$ on the simulation (scatter points) and analytical (curves) results. The parameter $\theta$ is set to 5 (circle), 10 (square), 30 (triangle), and 50 (diamond). (a)-(d) have the same meaning as do those in Fig. \ref{fig. 6}.}
		\label{fig. 9}
	\end{figure}

	Figure \ref{fig. 9} presents the simulation and analytical results as parameter $\theta$ increases from 5 to 50. As shown in Fig. \ref{fig. 9}(a), the analytical and simulation results show good agreement in terms of the tour length distribution. Moreover, we can observe that the probability of long tours remains the same as $\theta$ increases, indicating that $\theta$ has no influence on how individuals determine their tour length and the  probability of an individual’s termination. Moreover, as illustrated in Fig. \ref{fig. 9}(b), the growth rate of $S$ slightly increases as $\theta$ increases, since individuals have a greater likelihood of visiting new locations, reflecting an increase in individuals’ tendency to explore. Additionally, the analytical and simulation results exhibit noticeable deviations when the parameter $\theta$ is small, which can be attributed to the approximation steps involved in the analytical derivation (see Eq. (\ref{eq. 10})). However, the analytical results in Eq. (\ref{eq. 11}) are qualitatively consistent with the simulation results. Additionally, Fig. \ref{fig. 9}(c) demonstrates that the frequency at which individuals visit the base location remains the same, whereas the frequency of visits to nonbase locations becomes more homogeneous. This is because parameter $\theta$ cannot influence the probability of termination for an individual; thus, greater values of parameter $\theta$ reduce only the heterogeneity in the distribution of the frequency of visiting nonbase locations. Finally, Fig. \ref{fig. 9}(d) shows that, as the parameter $\theta$ increases, the distribution $P_{r_{g}}$ of the radius of gyration remains nearly unchanged, indicating that the parameter $\theta$ has little impact on $r_{g}$.
	
	\begin{figure}[h]
		\centering
		\includegraphics[width=0.48\textwidth]{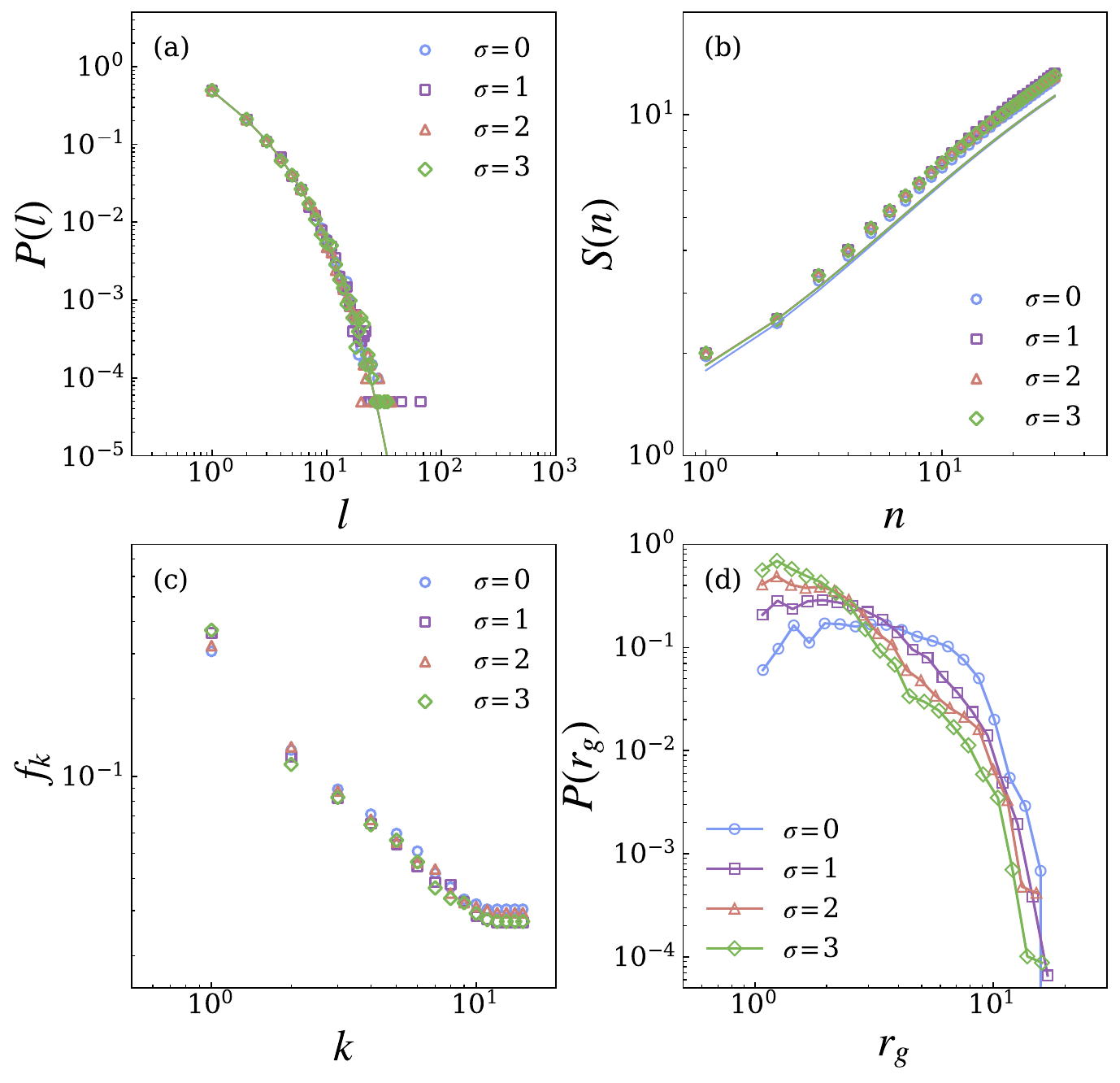}
		\caption{Effect of parameter $\sigma$ on the simulation (scatter points) and analytical (curves) results. The parameter $\theta$ is set to 5 (circle), 10 (square), 30 (triangle), and 50 (diamond). (a)-(d) have the same meaning as do those in Fig. \ref{fig. 6}.}
		\label{fig. 10}
	\end{figure}
	Figure \ref{fig. 10} presents the simulation and analytical results as the parameter $\sigma$ increases from 0 to 3. As shown in Fig. \ref{fig. 10}(a), the analytical and simulation results are in good agreement in terms of the tour length distribution. Moreover, we can observe that the probability of long tours remains the same as $\sigma$ increases, indicating that $\sigma$ has little influence on how individuals determine their tour length. As illustrated in Fig. \ref{fig. 10}(b), the growth rate of $S$ slightly remains the same, which indicates that $\sigma$ has little influence on how individuals explore new locations. Additionally, the analytical and simulation results exhibit noticeable deviations, which can be attributed to the approximation steps involved in the analytical derivation (see Eq. (\ref{eq. 10})). Furthermore, Fig. \ref{fig. 10}(c) demonstrates that the location visit frequencies remain the same. This findings indicates that the parameter $\sigma$ has little influence on the memory effect of human mobility. Finally, as shown in Fig. \ref{fig. 10}(d), as the parameter $\theta$ increases, the probability of low radius of gyration $r_{g}$ increase. As individuals favor closer destinations, their travel distances shorten, which decreases the overall radius of gyration $r_{g}$.

	\section{Discussion}
	In summary, we observed a universal scaling law of the tour length distribution in two cities. To our knowledge, this scaling law has been largely neglected in the literature on human mobility. To reproduce these characteristics, we propose a TTC model according to the empirical relationship between the probability of termination and the current tour length, as well as a competitive mechanism for choosing new or previously visited locations. We further extend the TTC model by a gravity-based exploration mechanism to capture the influence of spatial factors on individual tour behaviour. The extended model not only accurately reproduces the empirical tour length distributions but also successfully reproduces Heaps’ law, Zipf’s law and the distribution of the radius of gyration, offering a unified framework for understanding individual human mobility.
	
	Our study of individual tour behavior offers a foundation for both theoretical advancements and potential applications. In our work, the power law relationship between the probability of termination and the current tour length is the core mechanism of the TTC model. Analysis of the impact of changing the parameters on the tour length distribution indicates that to improve the efficiency of urban human mobility, it is necessary to either reduce the initial probability of termination or increase the degree to which the probability of termination decreases as the tour length increases (see Eq. (\ref{eq. 2})), thereby generating more long tours. For example, this can assist governments in formulating effective policies to guide citizens toward longer tours to reduce travel costs and increase the convenience of daily life in cities \cite{17}. Moreover, freight transportation companies can dynamically match freight tours to meet logistics needs by integrating real-time order demand with current truck locations \cite{33}. This approach not only reduces the number of trucks required but also alleviates environmental impacts such as greenhouse gas emissions. Additionally, by properly planning land use, it is possible to increase the accessibility and viability of urban services such as schools, shopping centers, healthcare facilities, and recreation so that citizens are more willing to conduct more activities within one tour \cite{34}. 
	
	While the proposed model effectively captures several key statistical characteristics of individual human mobility, it is not without limitations. First, our statistical analysis is based on Foursquare users' check-in data. Evaluating our hypothesis and model performance using mobility data from a broader range of data, such as mobile signaling data \cite{35} and private car GPS data \cite{36}, is important . Second, although the TTC model assumes a negative scaling relationship between the probability of tour termination and the current tour length, the underlying cause of this scaling relationship remains unclear. Subsequent explorations should aim to uncover the mechanisms that drive the observed decrease in the probability of termination as the tour length increases. Third, modeling the tour behavior of urban human mobility is only one application of the TTC model. The proposed tour-based mobility framework has the potential to be applied in several other domains that exhibit similar tour behaviors. For example, in freight transportation, trucks routinely visit multiple intermediate stops in a tour to pick up or deliver goods and then return to the base location. Applying the TTC model into this context could provide insights into tour behavior of trucks and assist in the design of more efficient logistic strategies (see Appendix B) \cite{14}. In terms of animal mobility, animals often travel within a limited range and regularly return to their home \cite{Borger2008}, which may be captured by the same tour-based mobility framework used in the TTC model. Building on our framework, the enhanced version could be used to help identify the conditions under which different foraging strategies emerge, or explain some observed animal mobility statistics. In all, further research may examine the TTC model's ability to simulate tour behavior in other systems, to better understand the complex mechanism of tour behavior.
	
	\section{Acknowledgement}
	This work was supported by the National Natural
	Science Foundation of China (Grant Nos. 72271019, 72288101). We thank Professor J. J. Ramasco for his insightful suggestions on our work.

	\appendix
	\section{Pseudocode for the TTC model}
	\FloatBarrier
	Note that $M$ is the total tour number of a city, $O$ is the number of visits of each location and $D$ is the distance matrix in a city. The base location is the location with the highest visit frequency for each truck.
	
	\begin{algorithm}[H]
		\caption{TTC model}
		\label{alg. 1}
		\begin{algorithmic}[1]
			\State \textbf{input:} $M$, $O$, $D$, \textit{base location}, $\rho$, $\gamma$, $\lambda$, $\sigma$
			\State \textbf{output:} \textit{trajectory}
			\State initialise a map \textit{frequencies} with all locations set to $0$
			\State set \textit{trajectory} $\gets$ [\textit{base location}], $l \gets 0$, \textit{current tour number} $\gets 0$, \textit{current location} $\gets$ \textit{base location}
			\While{\textit{current tour number} $< M$}
			\State $P \gets \Call{ReturnProb}{l, \rho, \gamma}$
			\State $u \gets \Call{RandUniform}{}$ \Comment{random variable $u \in (0,1)$}
			\State \textit{frequencies} $\gets$ \Call{UpdateFrequencies}{\textit{next location}, \textit{frequencies}}
			\If{$u < P$} \Comment{return to base}
			\State \textit{next location} $\gets$ \textit{base location}
			\State append \textit{next location} to \textit{trajectory}
			\State \textit{current tour number} $\gets$ \textit{current tour number} + 1
			\State \textit{current location} $\gets$ \textit{next location}
			\Else \Comment{continue: choose next location}
			\State \textit{next location} $\gets$ \Call{ChooseLoc}{\textit{frequencies}, \textit{current location}, $O$, $D$, $\lambda$, $\sigma$}
			\State \textit{frequencies} $\gets$ \Call{UpdateFrequencies}{\textit{next location}, \textit{frequencies}}
			\State append \textit{next location} to \textit{trajectory}
			\State $l \gets l + 1$;\quad \textit{current location} $\gets$ \textit{next location}
			\EndIf
			\EndWhile
			\State \Return \textit{trajectory}
		\end{algorithmic}
	\end{algorithm}

	\begin{algorithm}[H]
		\caption{Destination choice}
		\label{alg. 2}
		\begin{algorithmic}[1]
			\Procedure{ChooseLoc}{frequencies, current location, $O$, $D$, $\lambda$, $\sigma$}
			\State set a weight map \textit{w} $\gets$ copy of \textit{frequencies}
			\State add a special key \textit{newKey} with weight $\lambda$;\quad remove \textit{current location} from \textit{w}
			\State candidate $\gets$ \Call{WeightedRandom}{\textit{w}} \Comment{return a key proportionally to the provided \textit{weights}}
			\If{candidate equals \textit{newKey}}
			\State \textit{distancelist} $\gets$ the row of $D$ for \textit{current location}
			\State candidate $\gets$ \Call{GravityLaw}{\textit{frequencies}, \textit{current location}, $O$, \textit{distancelist}, $\sigma$}
			\EndIf
			\State \Return candidate
			\EndProcedure
			
			\Procedure{UpdateFrequencies}{next location, frequencies}
			\State increase \textit{frequencies}[next location] by $1$; \textbf{return}
			\EndProcedure
		\end{algorithmic}
	\end{algorithm}

	\begin{algorithm}[H]
		\caption{Return probability}
		\label{alg. 3}
		\begin{algorithmic}[1]
			\Procedure{ReturnProb}{$l$, $\rho$, $\gamma$}
			\If{$l == 0$}
			\State \Return $0$
			\Else
			\State \Return $\rho \cdot l^{-\gamma}$
			\EndIf
			\EndProcedure
		\end{algorithmic}
	\end{algorithm}
	
	\begin{algorithm}[H]
		\caption{Gravity law exploration}
		\label{alg. 4}
		\begin{algorithmic}[1]
			\Procedure{GravityLaw}{frequencies, current location, $O$, distancelist, $\sigma$}
			\State let $O_c$ be $O$ at index \textit{current location}
			\State compute score $s_j$ for each location $j$ using a gravity kernel
			\ForAll{location $j$ in \textit{frequencies} keys}
			\If{\textit{frequencies}[$j$] $> 0$} \State set $s_j \gets 0$ \Comment{forbid already-visited nodes} \EndIf
			\EndFor
			\State normalise $s$ to probabilities
			\State \textit{next location} $\gets$ \Call{WeightedRandom}{$s$} \Comment{return a key proportionally to the provided \textit{weights}}
			\State \Return \textit{next location}
			\EndProcedure
		\end{algorithmic}
	\end{algorithm}
	\FloatBarrier

	\section{Truck mobility}
	We apply TTC model in heavy truck mobility by using data in Beijing and Shanghai, which consists of intracity heavy truck GPS trajectories collected over a two-week period in 2018. Table \ref{tab. 2} presents the number of individuals, number of trips, number of tours, mean tour length and variance of tour length in the two cities.
	\label{truck}
	\begin{table}[h!]
		\centering
		\caption{Number of individuals, number of trips, number of tours, mean tour length, and variance of tour length in Shanghai and Beijing}
		\label{tab. 2}
		\begin{tabular}{lrr}
			\toprule
			\textbf{Metric} & \textbf{Shanghai} & \textbf{\quad Beijing} \\
			\midrule
			\textbf{Number of individuals} & 44{,}215 & 18{,}917 \\
			\textbf{Number of trips}       & 833{,}238 & 474{,}875  \\
			\textbf{Number of tours}       & 234{,}291 & 153{,}398  \\
			\textbf{Mean tour length}       & 13.23 & 16.72  \\
			\textbf{Variance of tour length}       & 131.25 & 189.14  \\
			\bottomrule
		\end{tabular}
	\end{table}

	\begin{figure}[h]
		\centering
		\includegraphics[width=0.5\textwidth]{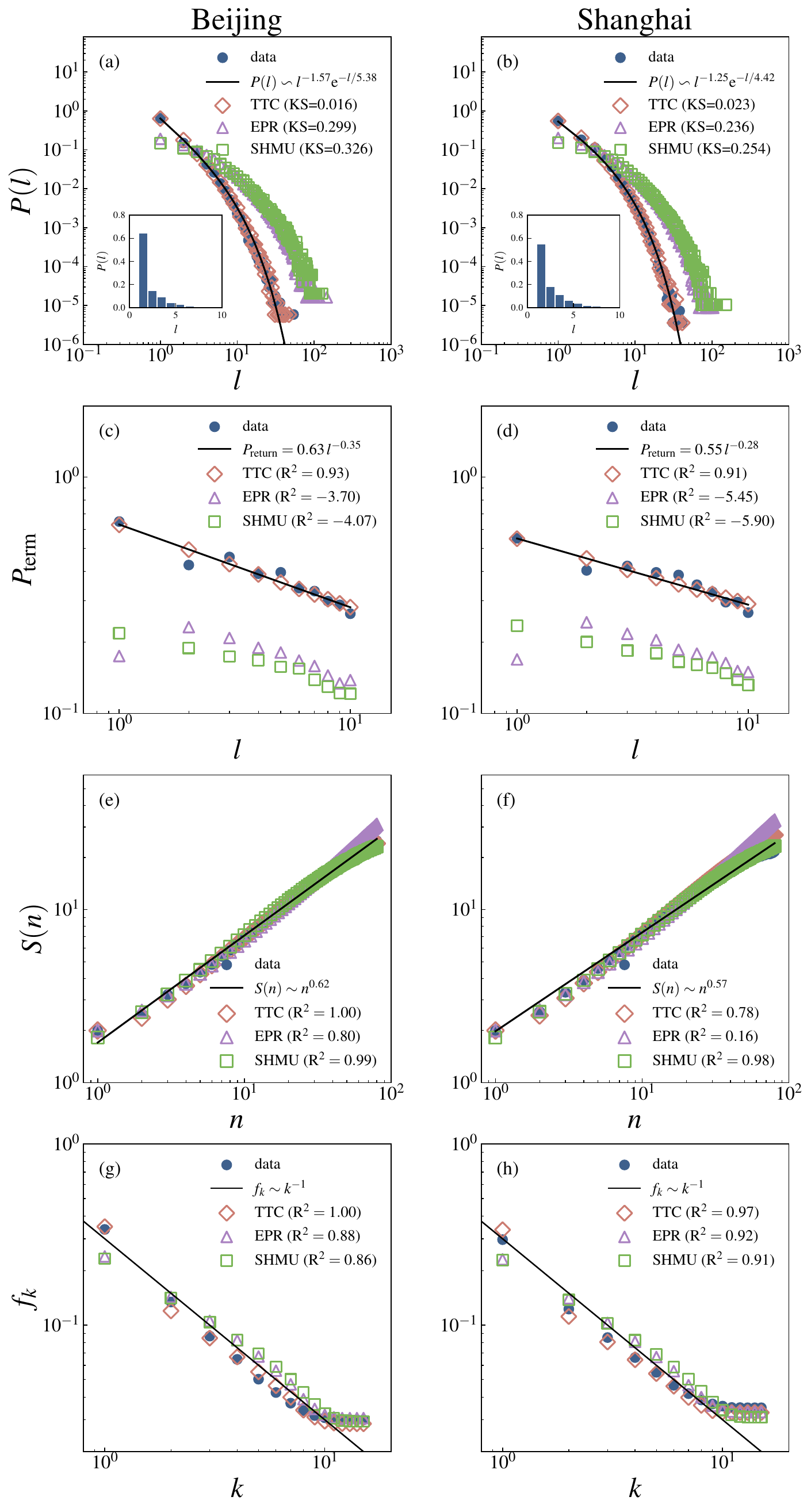} 
		\caption{Tour length distributions $P(l)$ of Beijing (a) and Shanghai (b). Scaling relationships between the probability $P_{\text{term}}$ of terminating the tour and the length $l$ of the current tour in Beijing (c) and Shanghai (d). Relationship between the number $S$ of locations visited and the number $n$ of trips in Beijing (e) and Shanghai (f). Distributions of the frequency of visits $f_{k}$ to locations in Beijing (g) and Shanghai (h). The meaning of the symbols and variables are consistent with those used in Fig. \ref{fig. 1}, Fig. \ref{fig. 3}, Fig. \ref{fig. 4} and Fig. \ref{fig. 5}.}
		\label{fig. 11}
	\end{figure}
		
	\FloatBarrier
	
	\bibliographystyle{unsrtnat}


\end{document}